# Doubly robust estimation of marginal cumulative incidence curves for competing risk analysis

Patrick van Hage[1,2], Saskia le Cessie[1,3], Marissa C. van Maaren[4,5], Hein Putter[1], Nan van Geloven[1]

Corresponding authors email: p.van.hage@biology.leidenuniv.nl

*[1] Department of Biomedical Data Sciences, Leiden University Medical Center, Leiden, the Netherlands*

*[2] Institute of Biology Leiden, Leiden University, Leiden, the Netherlands*

*[3] Department of Clinical Epidemiology, Leiden University Medical Center, Leiden, the Netherlands*

*[4] Department of Research and Development, Netherlands Comprehensive Cancer Organisation (IKNL), Utrecht, the Netherlands*

*[5] Department of Health Technology and Services Research, Technical Medical Centre, University of Twente, Enschede, the Netherlands*

## ABSTRACT
Covariate imbalance between treatment groups makes it difficult to compare cumulative incidence curves in competing risk analyses. In this paper we discuss different methods to estimate adjusted cumulative incidence curves including inverse probability of treatment weighting and outcome regression modeling. For these methods to work, correct specification of the propensity score model or outcome regression model, respectively, is needed. We introduce a new doubly robust estimator, which requires correct specification of only one of the two models. We conduct a simulation study to assess the performance of these three methods, including scenarios with model misspecification of the relationship between covariates and treatment and/or outcome. We illustrate their usage in a cohort study of breast cancer patients estimating covariate-adjusted marginal cumulative incidence curves for recurrence, second primary tumour development and death after undergoing mastectomy treatment or breast-conserving therapy. Our study points out the advantages and disadvantages of each covariate adjustment method when applied in competing risk analysis.

## 1. INTRODUCTION
Medical researchers are often interested in the effects of a treatment on the occurrence of a particular event in a setting where patients are at risk for more than one event. This may include the recurrence or progression of a disease, the occurrence of certain symptoms affecting the wellbeing of a patient, and death due to different causes.[1] Competing risk analysis refers to studying the time until occurrence of the first out of several possible events.

Failing to properly account for the risk of occurrence of other events may lead to an overestimation of the probability of occurrence of a particular event of interest over time, also referred to as cumulative incidence.[2]

In observational studies, when studying the effect of treatments we need to account for the fact that the choice of treatment observed in the data may have been influenced by the preferences of patients and clinicians.[3] The resulting imbalance of patient- or disease-related characteristics between treatment groups may confound the relationship between treatment and outcome. Differences in crude cumulative incidence curves will in such cases not reflect the causal effect of treatment. Estimating the effect of treatment from observational data requires adjustment for the imbalanced characteristics using causal inference techniques.[4]

Aiming for a fairer comparison between treatment groups, potential confounders are often included as covariates when modeling the relation between treatment and outcome.[5] In the setting of competing risks, this can for instance be done through cause-specific Cox regression models or with the Fine and Gray model.[2,6–8] Using these models it is possible to estimate adjusted conditional cumulative incidence curves, representing the cumulative incidence for an individual with given treatment and covariate values (e.g. an individual with their covariate values chosen at the mean or median of observed values). However, these curves are conditional on the covariate values and do not reflect the cumulative incidence in the entire patient population.
Several methods have been described to obtain marginal, i.e. population averaged, cumulative incidence curves which account for covariate imbalance between treatment groups.[9–11] Here we describe two popular methods to account for imbalance in baseline covariates: inverse probability weighting and outcome regression with standardization.
Inverse probability weighting is based on rebalancing the distribution of covariates between treatment groups using weighted cumulative incidence curves. To construct weighted estimators, the first step is modeling the probability of being assigned to a treatment group conditional on the set of confounders (i.e. the *propensity score*). Individual observations are then reweighted by the inverse of their propensity score.[12] Neumann et al. (2016) used this method within a competing risk analysis to adjust cumulative incidence curves for the imbalance in patient- and tumour-related covariates between two types of treatment for Hodgkin disease.[13] Furthermore, Choi et al. (2019) illustrated that this method may provide accurate marginal cumulative incidence estimates in unbalanced data.[14]
In the second method, *outcome regression* with standardization, the outcome is modelled as a function of both covariates and treatment using a competing risks analysis. Then, for each person, using their covariate values, two predicted cumulative incidence curves are calculated, one if the person would have been treated and one under no treatment. Marginal adjusted cumulative incidence curves per treatment group can then be obtained by averaging over the predicted curves under that particular treatment.[5,15] Outcome regression has been previously applied in a competing risks setting by Kipourou et al. (2019) to adjust for time-dependent covariates via a flexible regression model using cause-specific hazard models.[15,16]
In a later study by Hu et al. (2020), this method was applied using the Fine and Gray model to study the occurrence of treatment-related mortality and recurrence in leukaemia patients.[17]

Covariate adjustment methods generally rely on the assumptions of no unmeasured confounding, positivity, and consistency.[18] In addition, reweighting using propensity scores requires correct specification of the propensity score model, whereas the outcome regression approach requires correct specification of the outcome model. As a result, the robustness of these methods against various forms of model misspecification will differ. In this paper we propose a third, doubly robust, estimator, which combines the previous two methods. Doubly robust models require only one of these two models to be correctly specified.[19] Our proposal is an extension of the doubly robust approach of Wang (2018) for standard survival outcomes to the competing risks setting and makes use of pseudo-observations to deal with censoring.[20]

After verifying the doubly robust property theoretically, we compare the accuracy and precision of the proposed doubly robust method to that of inverse probability weighting method, the outcome regression method, and to the crude Aalen-Johansen estimator in a simulation study, including various scenarios of unbalanced treatment assignment and model misspecification. To demonstrate how these methods behave in a real-world data scenario, we apply them to a case study in breast cancer survival.

## 2. COVARIATE ADJUSTMENT METHODS

### 2.1 Cumulative incidence

We denote each individual by $i \in \{1, \dots, n\}$, with $n$ the total number of individuals in the dataset. We assume that there are $K$ possible events. Let $T$ be the time to the first event, and $C$ the censoring time. We assume $C$ to be independent of $T$, i.e. non-informative censoring. The observation time is defined as $T^* = \min(T, C)$, and $\Delta \in \{0, 1, \dots, K\}$ the event type that occurs first, with $\Delta = 0$ indicating the occurrence of censoring.

The cumulative incidence function for event $k$ is defined as $I_k(t) = P(T \leq t, \Delta = k)$. We can estimate the cumulative incidence function non-parametrically with the Aalen-Johansen estimator $\hat{I}_k(t)$, which is a function of the estimated total survival function $\hat{S}(t)$, representing the estimated probability to be free of any event at time $t$, and cause-specific hazard functions $\hat{\lambda}_k(t)$:

$$\hat{S}(t) = \prod_{j:t_j \leq t} \left(1 - \frac{d(t_j)}{n(t_j)}\right), \quad \hat{\lambda}_k(t_j) = \frac{d_k(t_j)}{n(t_j)},$$

$$\hat{I}_k(t) = \sum_{j:t_j \leq t} \hat{S}(t_j^-)\, \hat{\lambda}_k(t_j),$$

where $t_j \in \{0 < t_1 < \dots < t_j < \dots < t_J\}$ are the distinct event times where any of the $K$ competing events occur, $d_k(t_j) = \sum_i 1[T_i^* = t_j, \Delta_i = k]$ is the number of individuals failing from event $k$ at $t_j$, $d(t_j) = \sum_k d_k(t_j)$ is the total number of failures from any cause at $t_j$, and $n(t_j) = \sum_i 1[T_i^* \geq t_j]$ is the number of individuals who are still at risk at $t_j^-$.

### 2.2 Treatment specific cumulative incidence

We now consider the situation where we have two different treatment categories denoted by z=1 (treatment) and z=0 (control). Extension to more treatment categories is possible. We would like to estimate treatment specific marginal cumulative incidence curves defined as $I_k^{z=1}(t) = P(T^{z=1} \leq t, \Delta^{z=1} = k)$, the cumulative incidence over time had all individuals of the population received treatment, and $I_k^{z=0}(t) = P(T^{z=0} \leq t, \Delta^{z=0} = k)$, the cumulative incidence function had all individuals not received treatment. The superscript notation indicates that we aim for estimating the distribution of potential outcomes, that is outcomes for individuals under treatment categories that may be different than the ones actually observed in the data.

When data from a perfectly randomized trial would be available, the treatment specific cumulative incidence curves could be estimated using the non-parametric Aalen-Johansen estimator described in Section 2.1 applied to the observed data from each treatment group. However, when treatment is not assigned randomly, results are prone to bias due to confounding and covariate adjustment methods are needed.

To describe different covariate adjustment methods for marginal cumulative incidence functions, we consider an observational data set with $Z_i \in \{0, 1\}$ the received treatment of individual i and with $X_i$ a vector of confounders known at baseline. We assume that there is no unmeasured confounding and that other identifying assumptions for causality (positivity and consistency) hold.[4] More formal definitions of these assumptions are given in Appendix A.

## 2.3 Inverse probability weighting

With covariate adjustment via inverse probability weighting, we model the treatment allocation as a function of the confounders. This information is then used to rebalance the treatment groups via weighting such that the distribution of covariates becomes similar in both weighted treatment groups.[12,21] To this end, the propensity score is used, which is the probability of an individual $i$ being assigned treatment $Z_i = 1$, given its covariate values $X_i$.[12,13,21] The propensity score can be estimated through any classification-based method, but is most commonly estimated through logistic regression:[12]

$$P(Z_i = 1|X_i) = \frac{\exp(\gamma_0 + \gamma X_i)}{1 + \exp(\gamma_0 + \gamma X_i)}.$$

The corresponding inverse probability weights are calculated based on the observed treatment values $Z_i$:

$$w_i = \frac{1}{Z_i P(Z_i = 1|X_i) + (1 - Z_i)\big(1 - P(Z_i = 1|X_i)\big)}$$

As such, individuals with covariate values which are overrepresented in a treatment group receive a smaller weight, and vice versa. This yields reweighted numbers of events of type $k$ at $t_j$, total number of events at $t_j$, and numbers at risk at $t_j$, in both treatment groups expressed as:

$$\hat{d}_{IPW,k}^{z}(t_j) = \sum_i w_i \, 1\big[T_i^* = t_j, Z_i = z, \Delta_i = k\big],$$

$$\hat{d}^z_{IPW}(t_j) = \sum_k \hat{d}^z_{IPW,k}(t_j)$$

$$\hat{n}^z_{IPW}(t_j) = \sum_i w_i \, 1[T^*_i \geq t_j, Z_i = z].$$

The weighted estimators for the overall survival function and cause-specific hazard functions become:

$$\hat{S}^z_{IPW}(t) = \prod_{j:t_j \leq t}\left(1 - \frac{\hat{d}^z_{IPW,k}(t_j)}{\hat{n}^z_{IPW}(t_j)}\right), \quad \hat{\lambda}^z_{IPW,k}(t_j) = \frac{\hat{d}^z_{IPW,k}(t_j)}{\hat{n}^z_{IPW}(t_j)}.$$

Finally, combining these estimators yields the reweighted cumulative incidence estimator under treatment category $z$:

$$\hat{I}^z_{IPW,k}(t) = \sum_{j:t_j \leq t} \hat{S}^z_{IPW}(t_j^-) \, \hat{\lambda}^z_{IPW,k}(t_j).$$

Point wise standard errors for $\hat{I}^z_{IPW,k}(t)$ may be approximated analytically.[14] However, these methods do not account for the variation in the estimation of the propensity scores. Bootstrap methods can be used instead.[22]

## 2.4 Outcome regression with standardization

In this approach we use that the treatment specific marginal cumulative incidence functions can be rewritten as

$$I^{z=1}_k(t) = \int I^{z=1}_k(t|x) f(x) dx,$$
$$I^{z=0}_k(t) = \int I^{z=0}_k(t|x) f(x) dx.$$

with integration over the covariate distribution of the population. In the outcome regression approach, $I^z_k(t|x)$ is obtained through regression modelling. In this paper we use cause-specific Cox proportional hazards models, modelling the effect of treatment and covariates, $Z$ and $X$, on the cause-specific hazards, but other competing risks regression models may be used as well, such as the Fine and Gray model. Integration over the covariate distribution can be approximated by averaging over the observed covariate values, i.e. the empirical distribution, as follows

$$\hat{I}^z_{OR,k}(t) = \frac{1}{n}\sum_{i=1}^{n} \hat{I}_k(t, X_i, Z = z).$$

In this method, the regression model is used to predict two cumulative incidence curves for every individual given their covariate values, once setting $Z = 1$ and once $Z = 0$. Then the average is taken over all individuals to obtain the estimated marginal cumulative incidence curves for each treatment. Standard errors can be obtained via bootstrapping or analytically (e.g. as described by Sjölander, 2020).[23]

## 2.5 Doubly robust estimation

The previous two covariate adjustment methods have different underlying assumptions. Inverse probability weighting requires correct specification of the propensity score model, whereas outcome regression requires correct specification of the outcome model. We now propose a third method using an augmented IPW estimator. Augmented IPW estimators are constructed as a weighted sum of the IPW estimator and outcome regression model. The general form of the augmented IPW estimator for the treatment specific mean of a continuous variable $Y$ under treatment category $z$, i.e., $E(Y^z), is$:

$$\hat{Y}_{DR}^{z} = \frac{1}{n}\sum_{i=1}^{n}\left[\frac{Y_i\,1[Z_i = z] - \hat{Y}(X_i, Z = z)\left(1[Z_i = z] - \hat{P}(Z_i = z|X_i)\right)}{\hat{P}(Z_i = z|X_i)}\right],$$

with $\hat{P}(Z_i = z|X_i)$ the estimated propensity score and $\hat{Y}(X_i, Z = z)$ an estimate of the outcome regression model. One can see from the formula that if the propensity score model fits the data well, i.e., $1[Z_i = z] - \hat{P}(Z_i = z|X_i)$ is close to zero, and the estimator reduces to a regular IPW estimator. Similarly one show that if the estimates from the outcome model $\hat{Y}(X_i, Z = z)$ are close to the observed outcomes $Y_i$ in individuals with observed $Z_i = z$, the estimator reduces to the outcome regression estimator. This results in the so called double-robustness property: as long as either the propensity score model or the outcome regression model is correctly specified, the estimator is consistent.[24–26] A simulation study by Lunceford et al. (2014) demonstrated that this augmentation also makes the estimator generally more efficient, i.e. having a smaller mean-squared error, when compared to the propensity score weighted estimator.[24] Though Lunceford et al. noted that its efficiency does fall short when compared to a correctly specified outcome model approach.

Defining the augmented IPW estimator in the context of survival analysis is challenging, as the outcome, i.e., the time until event, is not observed for all individuals due to censoring. An augmented IPW estimator for the survival probability at time $t$ in the ordinary survival setting was proposed by Wang (2018), where the observed status indicator by time $t$ used in the doubly robust estimator was replaced by the pseudo-value for the survival function at that time, derived from the Kaplan Meier estimator. We extend this approach to the setting with competing risks. Pseudo-values that represent an individual's contribution to the cause-specific incidence function have been formulated by Andersen & Perme.[27,28]

$$y_{i,k}^{*}(t) = n \cdot \hat{I}_k(t) - (n-1) \cdot \hat{I}_k^{-i}(t),$$

with $\hat{I}_k^{-i}(t)$ the cumulative incidence by time $t$, estimated by the Aalen-Johansen estimator, obtained leaving out the $i$th observation. Our proposed doubly robust estimator of the treatment specific marginal cumulative incidence function at time $t$ can then be expressed as

$$\hat{I}_{DR,k}^{z}(t) = \frac{1}{n}\sum_{i=1}^{n}\frac{y_{i,k}^{*}(t)\,1[Z_i = z] - \hat{I}_k(t, X_i, Z_i = z)\left(1[Z_i = z] - \hat{P}(Z_i = z|X_i)\right)}{\hat{P}(Z_i = z|X_i)}$$

A derivation of the consistency of this estimator under correct specification of either the outcome model or the propensity model, is presented in Appendix A.

## 3. SIMULATION STUDY

## 3.1 Simulation set-up

### *3.1.1 Data generation*

A simulation study was conducted to evaluate the performance of the three methods, and their robustness against misspecification of the association between the covariates and treatment assignment and/or outcome. We simulated $N$ individuals, $i \in \{1, \ldots, N\}$. For each individual we simulated two independent covariates: one categorical variable with three categories drawn from a multinomial distribution with each category having an equal 1/3 probability, represented by two indicator variables $X_1, X_2$, plus a continuous covariate $X_3 \sim \mathrm{N}(0, 1)$.

We considered two competing events $k \in \{1,2\}$. In each simulation scenario we assumed a proportional hazards model for the cause specific hazard of each event:

$$\lambda_k(Z_i, X_i) = \exp(\lambda_{k,0}) \cdot \exp(X_i \beta_k + Z_i \theta_k),$$

with $\exp(\lambda_{k,0})$ a constant baseline hazard, $\theta_k$ the event-specific treatment effect on the log hazard, and $\beta_k$ the vector of covariate effects on the log hazard. The individual (possibly latent) event times for each event type $k$, $T_{ik}$ were generated as described by Bender et al.[29],

$$T_{ik} = \frac{\log(U_{ik})}{\lambda_k(Z_i, X_i)},$$

where $U_{ik}$ was drawn from a uniform distribution $U_k \sim U(0, 1)$. The time of the first event was derived as the minimum of the event times for the two event types: $T_i = min(T_{i1}, T_{i2})$. Censoring times $C_i$ were sampled from a uniform distribution: $C_i \sim \mathrm{U}(P_{.20}(T), P_{.95}(T))$, where $P_{.20}(T)$ and $P_{.95}(T)$ were calculated for each simulation run as the 20th and 95th percentile from the empirical distribution of the event times $T$. This resulted in roughly 25% of censored observations under all simulation conditions. The observation time was $T_i^* = \min(T_i, C_i)$, and the observed status $\delta_i$ was set to 0 if $T_i > C_i$, $\delta_i = 1$ if $T_{i1} < T_{i2}$ and $T_i < C_i$, and $\delta_i = 2$ if $T_{i1} > T_{i2}$ and $T_i < C_i$.

### *3.1.2 Simulation scenarios*

We generated four main scenarios: one scenario leading to correctly specified outcome and exposure models in the analysis (Scenario 1), one which led to misspecification in the exposure model (Scenario 2), one with misspecified outcome model (Scenario 3) and one with both models misspecified (Scenario 4). We used the following parameter values in all four scenarios: treatment effects $\theta_1 = -1$, $\theta_2 = -0.5$, coefficients for the covariates for the first event type $\beta_1 = (1, -1, 0.5)$, and coefficients for the second event type $\beta_2 = (-1, 1, -0.5)$. In Scenario 1 and Scenario 2, we chose baseline hazard $\exp(\lambda_{k,0}) = 1$ for both event types. These parameter choices ensured that the positivity assumption was met in each of the four scenarios. In supplementary Section S2 we show performance of the methods in an additional scenario where the positivity assumption is violated.

**Scenario 1: Correct specification of both the treatment and the outcome model.** A logistic regression model was used to generate the probability of being assigned the active treatment $(Z_i = 1)$, as a function of the covariate values $X_i$:

$$Pr(Z_i = 1 | X_i) = \frac{\exp(X_i \, \omega)}{1 + \exp(X_i \, \omega)},$$

with coefficients $\omega = (1, -1, 1)$.

**Scenario 2: Misspecification of the treatment model.** Next, we generated data from a scenario where individuals were assigned to treatment in a way that does not align with a logistic treatment model. We first generated data under Scenario 1 and only kept individuals with $Z = 1$ ($n_1$ individuals). Next we generated a large dataset where each individual was randomly assigned one of two treatments, $Z_i \sim \mathrm{Bern}(0.5)$ and selected $(N - n_1)$ individuals from this large dataset with $Z = 0$. Consequently, in the combined dataset the relationship between covariates and treatment allocation no longer follows a logistic regression model.

**Scenario 3: Misspecification of the outcome model.** In this simulation scenario, the relationship between covariates and outcomes is misspecified, by generating a non-linear relationship between the log hazard and the continuous covariate $X_3$. This was done by generating latent survival times for each cause as follows:

$$\lambda_k(Z_i, X_i) = \exp(\lambda_{\mathrm{k},0}) \cdot \exp(Z_i \theta_k + X_i \beta_k - 4[X_{3,i} < 1])$$

with $\lambda_{k,0} = 2$ for both event types. The treatment assignment model was the same as in Scenario 1.

**Scenario 4: Misspecification of both the treatment and the outcome model.** In this simulation scenario, the data is generated under the conditions described both by Scenario 2 and Scenario 3. Consequently, the relationship between covariates and treatment allocation does not follow a logistic model, and the relationship between the log hazard and continuous covariate $X_3$ is non-linear.

## 3.2 Performance evaluation

Simulations for each scenario were repeated $M = 1000$ times. In each scenario the sample size was $N = 4000$. The accuracy and precision of the crude estimator of the marginal cumulative incidence curves, i.e. applying the unadjusted Nelson Aalen estimator in each treatment group, and of the three covariate adjustment methods were evaluated. We expressed performance in terms of bias, i.e. the mean difference between the estimated marginal cumulative incidence and the underlying true cumulative incidence, per time point and in terms of root mean squared error (RSME), i.e. the root of the mean squared difference between estimated and true cumulative incidences per time point. Derivation of the true incidence can be found in supplementary Section S1.[30] The bias with pointwise 2.5th and 97.5$^{\text{th}}$ percentiles of the differences derived from the simulations, was plotted over a range of time points. We also report bias and RMSE at a time point approximately halfway during follow up.

## 3.3 Statistical software and packages

R statistical software (version 4.1.3) was used. The covariate adjustment methods were implemented using the package *survival* for the inverse probability weighting method, *riskRegression* for the outcome regression and doubly robust method, and *prodlim* to generate pseudo-observations for competing risk.[27,31,32] Visualizations were produced with the *ggplot2* and *firatheme* package.[33,34] Performance measures and diagnostic procedures were derived using the *microbenchmark, survey,* and *tableone* packages.[35–37] The R-code for the covariate adjustment methods and for the simulation study is available on https://github.com/survival-lumc/AdjCuminc.

## 3.4 Results

Figure 1 depicts the deviation from the true marginal cumulative incidence curve for each estimation method in Scenarios 1 to 4, plotted over the range of simulated event times. The results from Scenario 1 (correct treatment and outcome model specification) illustrate substiantial bias in the naïve estimator (Figure 1A). The bias in all three covariate adjusted estimators was negligible. The RMSE of the outcome regression method was somewhat smaller compared to the propensity score-based method and the doubly robust method (Table 1). In Scenario 2 (Figure 1B), where the treatment model was misspecified, the inverse probability weighting method yielded biased estimates. In contrast, both outcome regression and the doubly robust method were unbiased, with equal or smaller RMSE for the outcome regression method. The results for the third scenario (Figure 1C), with a misspecified outcome model, show that the incidence curves using outcome regression were biased while the other two methods were not. Even though, the RMSE was still in some situations lower for the outcome regression method than that of the other two methods (e.g., for event type 1 in the control group and for event type 2 in the treatment group at $t = 5.0$, Table 1). Lastly, we observe that none of the methods perform well in Scenario 4, when both the treatment allocation and outcome model were misspecified. The doubly robust method did not perform worse compared to the other two methods in this scenario.

To summarize, the doubly robust estimator was found to be robust against misspecification of either the treatment allocation or outcome model. When both models were misspecified, the doubly robust estimator did not perform better, but also not worse, than the other methods. Outcome regression was demonstrated to be the most efficient estimator when the outcome model was correctly specified.

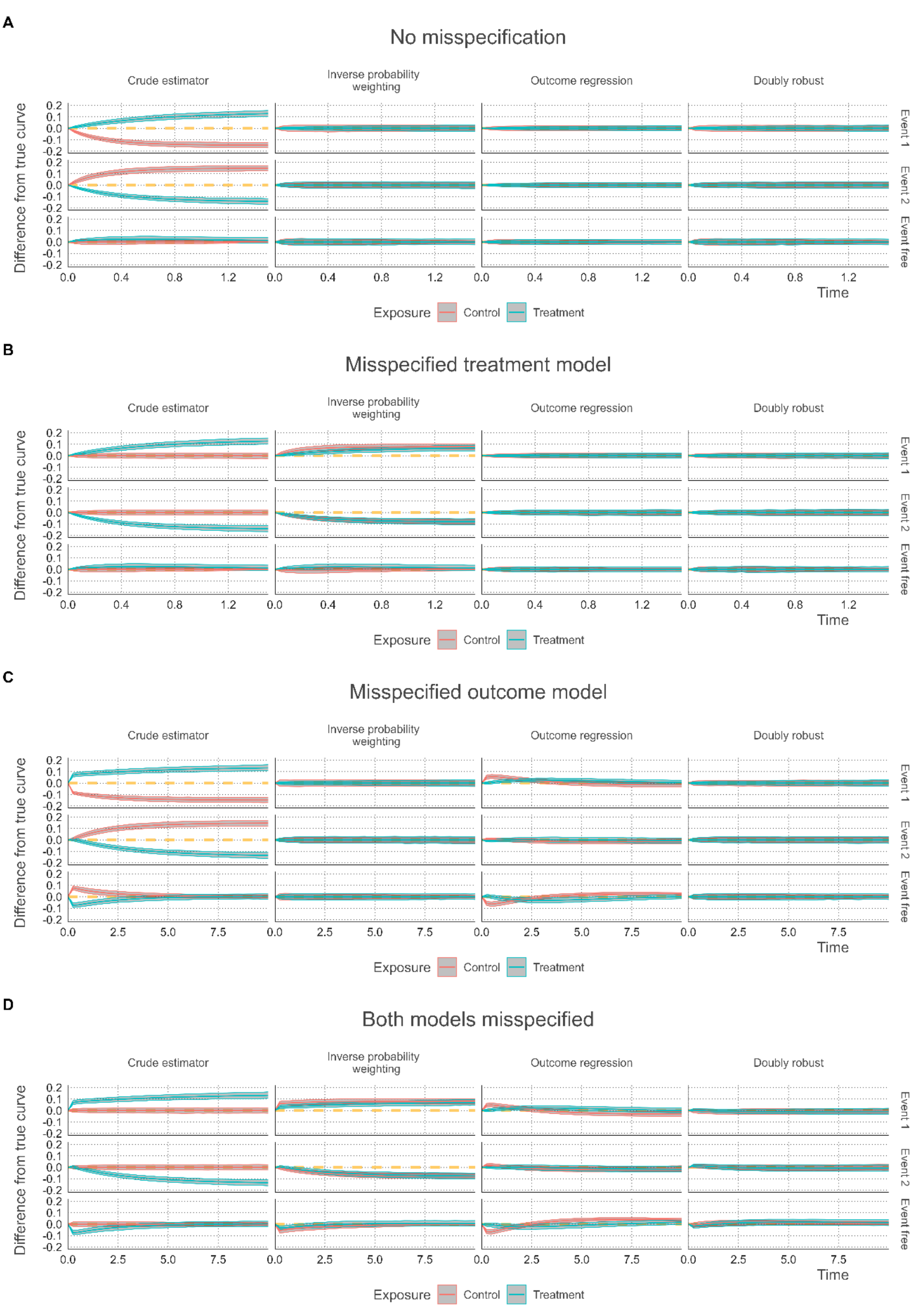

**Figure 1.** Simulation results for Scenario 1, no model misspecification (A), Scenario 2, misspecified covariate effects on treatment allocation (B), Scenario 3, misspecified covariate effects on the outcome (C), and Scenario 4, both treatment allocation and covariate effects on the outcome misspecified (D). Mean difference between the true cumulative incidence of event 1, 2, and event free survival and the estimated results using the naïve estimator (no covariate adjustment), inverse probability weighted estimator, outcome regression estimator, and doubly robust estimator in the two treatment groups. The bands indicate the 2.5$^{th}$ and 97.5$^{th}$ percentiles.

**Table 1. Simulation results under four scenarios:** Scenario 1: no model misspecification, Scenario 2: misspecification of covariate effects on treatment allocation, Scenario 3 misspecification of covariate effects on the outcome, and Scenario 4, both treatment allocation and covariate effects on the outcome misspecified, at approximately halfway the follow up. Bias and root-mean squared error (RMSE) of the naïve estimator (no covariate adjustment), inverse probability weighted estimator, outcome regression estimator, and doubly robust estimator of the cumulative incidence are given for the cumulative incidence of event 1 and event 2, and event free survival.

| **1. No misspecification (t = 0.8)** | | **CI event 1** | | **CI event 2** | | **Event free survival** | |
|---|---|---|---|---|---|---|---|
| | | **Bias** | **RMSE** | **Bias** | **RMSE** | **Bias** | **RMSE** |
| **Crude estimator** | **Control** | -0.14006 | 0.14049 | 0.14016 | 0.14067 | -0.00010 | 0.00844 |
| | **Treatment** | 0.10121 | 0.10189 | -0.12433 | 0.12481 | 0.02312 | 0.02604 |
| **Inverse probability** | **Control** | -0.00039 | 0.01431 | 0.00027 | 0.01306 | 0.00012 | 0.00973 |
| | **Treatment** | 0.00029 | 0.01062 | -0.00053 | 0.01502 | 0.00025 | 0.01352 |
| **Outcome regression** | **Control** | 0.00026 | 0.01055 | -0.00020 | 0.00992 | -0.00006 | 0.00757 |
| | **Treatment** | 0.00042 | 0.00876 | -0.00055 | 0.01142 | 0.00013 | 0.01115 |
| **Doubly robust** | **Control** | 0.00017 | 0.01296 | -0.00016 | 0.01208 | -0.00001 | 0.00968 |
| | **Treatment** | 0.00024 | 0.01035 | -0.00049 | 0.01361 | 0.00025 | 0.01311 |

| **2. Misspecified treatment (t = 0.8)** | | **CI event 1** | | **CI event 2** | | **Event free survival** | |
|---|---|---|---|---|---|---|---|
| | | **Bias** | **RMSE** | **Bias** | **RMSE** | **Bias** | **RMSE** |
| **Crude estimator** | Control | 0.00004 | 0.01186 | -0.00023 | 0.01158 | 0.00019 | 0.00806 |
| | Treatment | 0.10009 | 0.10075 | -0.12325 | 0.12372 | 0.02316 | 0.02596 |
| **Inverse probability** | Control | 0.07741 | 0.07827 | -0.07335 | 0.07410 | -0.00406 | 0.00900 |
| | Treatment | 0.05193 | 0.05297 | -0.07270 | 0.07359 | 0.02076 | 0.02399 |
| **Outcome regression** | Control | 0.00005 | 0.01085 | -0.00025 | 0.01056 | 0.00020 | 0.00683 |
| | Treatment | -0.00011 | 0.00911 | -0.00004 | 0.01219 | 0.00015 | 0.01167 |
| **Doubly robust** | Control | 0.00021 | 0.01220 | -0.00037 | 0.01171 | 0.00016 | 0.00822 |
| | Treatment | 0.00003 | 0.01007 | -0.00006 | 0.01281 | 0.00003 | 0.01233 |

| **3. Misspecified outcome (t = 5.0)** | | **CI event 1** | | **CI event 2** | | **Event free survival** | |
|---|---|---|---|---|---|---|---|
| | | **Bias** | **RMSE** | **Bias** | **RMSE** | **Bias** | **RMSE** |
| **Crude estimator** | **Control** | -0.14269 | 0.14309 | 0.13332 | 0.13385 | 0.00938 | 0.01301 |
| | **Treatment** | 0.11556 | 0.11613 | -0.10810 | 0.10862 | -0.00746 | 0.01359 |
| **Inverse probability** | **Control** | -0.00056 | 0.01379 | 0.00044 | 0.01278 | 0.00012 | 0.00931 |
| | **Treatment** | 0.00025 | 0.01031 | -0.00004 | 0.01480 | -0.00020 | 0.01369 |
| **Outcome regression** | **Control** | -0.00267 | 0.00992 | -0.01609 | 0.01874 | 0.01876 | 0.02005 |
| | **Treatment** | 0.02329 | 0.02475 | 0.00161 | 0.01132 | -0.02490 | 0.02711 |
| **Doubly robust** | **Control** | 0.00012 | 0.01244 | -0.00002 | 0.01187 | -0.00010 | 0.00927 |
| | **Treatment** | 0.00017 | 0.01013 | -0.00002 | 0.01353 | -0.00015 | 0.01340 |

CI: cumulative incidence; RMSE: root mean squared error

| 4. Both misspecified (t = 5.0) | | CI event 1 | | CI event 2 | | Event free survival | |
|---|---|---|---|---|---|---|---|
| | | Bias | RMSE | Bias | RMSE | Bias | RMSE |
| **Crude estimator** | **Control** | 0.00002 | 0.01185 | -0.00019 | 0.01165 | 0.00017 | 0.00868 |
| | **Treatment** | 0.11470 | 0.11529 | -0.10733 | 0.10790 | -0.00738 | 0.01365 |
| **Inverse probability** | **Control** | 0.07937 | 0.08018 | -0.07005 | 0.07084 | -0.00931 | 0.01243 |
| | **Treatment** | 0.05789 | 0.05882 | -0.06275 | 0.06385 | 0.00486 | 0.01297 |
| **Outcome regression** | **Control** | -0.01986 | 0.02230 | -0.01774 | 0.02061 | 0.03760 | 0.03828 |
| | **Treatment** | 0.01760 | 0.01991 | -0.00825 | 0.01477 | -0.00935 | 0.01466 |
| **Doubly robust** | **Control** | -0.00780 | 0.01454 | -0.00277 | 0.01207 | 0.01057 | 0.01383 |
| | **Treatment** | -0.00919 | 0.01395 | -0.00720 | 0.01524 | 0.01639 | 0.02094 |

CI: cumulative incidence; RMSE: root mean squared error

# 4. APPLICATION TO BREAST CANCER SURVIVAL DATA

## 4.1 Data characteristics

The three covariate adjustment methods were applied to breast cancer survival data from a previous cohort study by van Maaren et al.[38] The data were collected in collaboration with the Netherlands Comprehensive Cancer Organisation, and consisted of 8879 patients from the Netherlands Cancer Registry of whom we selected 6538 patients who underwent either a mastectomy (i.e. surgical removal of the entire breast) or breast-conserving therapy, and had complete information on all relevant covariates. Details on data collection, the study population, and classification and staging of cancer can be found in van Maaren et al. (2019) and supplementary Section S3.[38] An explorative study by van Maaren et al. (2016) found patients who received breast-conserving therapy had at least equal survival compared to patients who received a mastectomy.[39] The 2019 study compared different methods to correct for confounding, concluding that propensity score weighting, instrumental variable analysis and multivariable regression modeling all led to similar results and that assumptions of these methods have to be very carefully considered.[38] For example, unmeasured confounding could be present. Multiple event types could occur during follow up: locoregional and distant recurrences, second primary tumour development, and death without any such event. In the previous studies, the endpoints overall survival and 'distant metastasis free survival' (censoring observations if other events occurred) were used. In our analysis, we study recurrence (including locoregional and distant recurrences), second primary breast tumours, and death as separate competing events. The aim is to shed more light on the role of confounding by studying the adjusted cumulative incidence curves for each event type separately.

A total of 4116 patients (63.0%) received breast-conserving therapy, whereas 2422 patients (37.0%) underwent mastectomy. Breast-conserving therapy is recommended for early-stage disease when medically feasible, and according to patient preferences.[40,41] Patients in earlier stages (eg. smaller tumour size, fewer positive lymph nodes, no multifocality, ductal histological features, etc.) more frequently received breast-conserving therapy *(*see supplementary Table S2). All covariates that were previously included in van Maaren et al. (2019) were used for covariate adjustment, excluding TNM staging classifiers due high collinearity with tumour characteristics (tumour size, number of positive lymph nodes, and tumour grade). Of the sixteen patient-, hospital-, and tumour-related covariates, eight differed significantly between treatment groups. As such, comparison of crude cumulative incidence curves between treatment groups is likely to be confounded.

## 4.2 Comparison of covariate adjusted cumulative incidence curves

To estimate the propensity scores, treatment allocation was modeled as function of the covariates using logistic regression (see supplementary Figure S1 for the distribution of the propensity scores). The coefficients for the propensity score model can be found in supplementary Table S2. Considerable improvement on covariate balance was found after adjustment, with differences between covariates all within the 0.1 SEM threshold after propensity score weighting (see supplementary Figure S1). For the outcome regression approach regression coefficients of the Cox cause-specific hazard regression models can be found in supplementary Table S3).

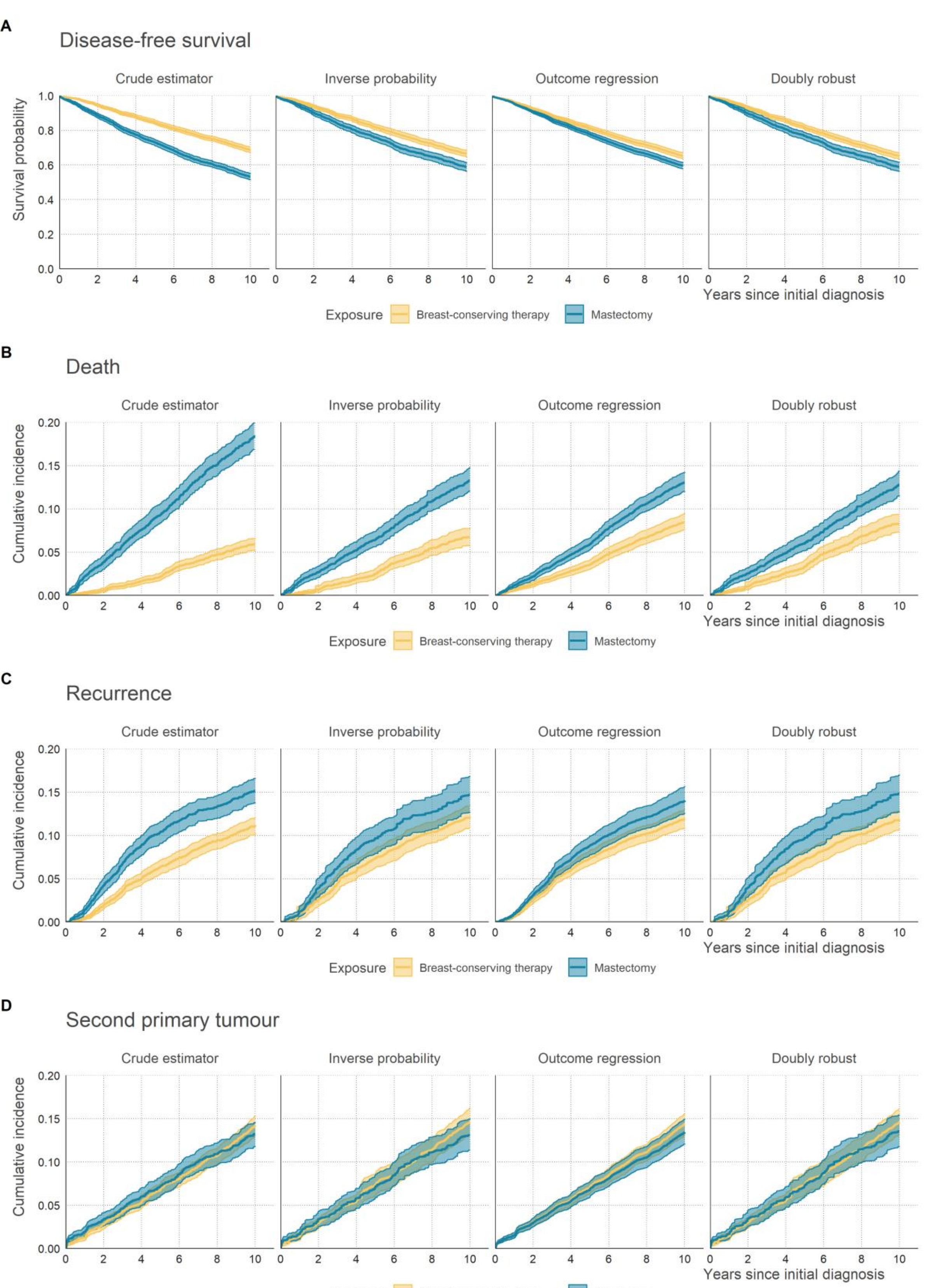

**Figure 2.** Adjusted cumulative incidence curves for (A) disease-free survival, i.e. the probability of not experiencing any of the event types, (B) death, (C) recurrence, and (D) second primary tumour development for breast-conserving therapy (yellow; light) or mastectomy (blue; dark). The cumulative incidence is shown for the naïve cumulative incidence estimator, inverse probability weighted estimator, outcome regression estimator, and doubly robust estimator. Confidence intervals were obtained by bootstrapping (B = 250 resamples).

We checked for deviations of the proportional hazard assumption by testing for independence between scaled Schoenfield residuals and time. For some of the covariates in the model for recurrence, the assumption of proportionality was violated (Tumour size, Grade, Hormonal receptor status, and Systemic therapy), which may be indicative of a lack of fit in this particular outcome model. This could be addressed by e.g. allowing time-varying coefficients, but this was not further explored here. Linearity of the three continuous covariates was assessed by plotting martingale residuals, no relevant deviations were observed (Supplementary Figure S3).

The estimated covariate adjusted cumulative incidence curves for each of the three event types are shown in Figure 2. The largest effects of covariate adjustment are found on the cumulative incidence of death (Figure 2B). The results were similar across the three adjustment methods, showing a lower cumulative incidence of death for patients that underwent mastectomy, and a higher cumulative incidence of death for patients that received breast-conserving therapy compared to the crude estimator. In addition, confidence bounds showed more overlap between treatment groups after adjustment (Figure 2C). Confidence bounds were wider after covariate adjustment, notably for the propensity-score method and the doubly robust method. The confidence intervals for the treatment groups largely overlapped for secondary tumour development, both with and without covariate adjustment (Figure 2D). Based on these observations, it can be said that the imbalance in the distribution of these recorded covariates mainly impacted results for patient mortality, and to some extent recurrence of the disease. However, it should be reiterated that the results described here are likely prone to unmeasured confounding , which has also been discussed before by van Maaren et al. (2019), and should therefore not be used to support clinical decision making.

## 5. DISCUSSION

In this paper we introduced a doubly robust method for estimating covariate-adjusted marginal cumulative incidence curves using pseudo-observations and compared its accuracy and precision to that of inverse probability weighting and outcome regression. The simulation study confirmed the doubly robustness of our newly proposed estimator, it yielded unbiased results if one of the two models (for treatment or outcome) was correctly specified. In addition, when both models were misspecified, the doubly robust estimator performed similar to the inverse probability weighted estimator and outcome regression estimator. The doubly robust estimator is therefore recommended if there is uncertainty on which model can be correctly specified.

A disadvantage of the doubly robust estimator is its relatively high variance compared to the outcome regression estimator.[42] The variance was similar to the variance of the inverse probability weighting method, which indicates that the use of pseudo-observations did not additionally increase the variance. This finding is similar to what was found earlier for linear outcome models and ordinary survival analysis.[5,20,26,42,43] Therefore, outcome regression is recommended, if the researcher has confidence in the outcome model. The outcome model estimate is in general also less influenced by situations where there is near non-positivity. A third limitation of the proposed doubly robust method is that the estimate for each time point is generated independently of the others, and may therefore not guarantee a monotonous curve.[26] However, the curves showed negligible non-monotonous deviation when the estimator was applied on the simulated data during preliminary trial runs, and when the methods were applied on real-world cancer survival data.

Another consideration is the role of censoring on the performance of these methods. In our simulation study, we have only considered a non-informative censoring mechanism. For the outcome regression method, this assumption can be relaxed to independent censoring conditional on the covariates. A caveat of the doubly robust estimator is that the modeling of pseudo-observations requires independent censoring.[28] Therefore, it can be expected that the proposed doubly robust estimator is sensitive to dependent censoring. Extensions of pseudo-observations to settings with dependent censoring have been proposed,[44,45] but using them in the current context has yet to be further explored. It should be noted that simulation studies are never exhaustive. The simulation code has been made publicly available, and can be readily adapted in follow-up studies to more specifically examine the behavior of these estimators under different types of censoring and using different parameters for the outcome and treatment models.

Alternative augmented IPW estimators for causal analysis of censored data have been proposed in literature that, instead of pseudo-observations, use inverse probability of censoring weighting to account for censoring,[46–48] including the setting of competing events.[49,50] An overview of methods focusing on standard survival data is given in Denz et al. (2023) and in Gabriel et al (2024).[26,51]

We used bootstrapping methods to generate confidence intervals in the breast cancer application. An analytical estimator for the variance of the doubly robust method for ordinary survival was proposed by Wang (2018),[16] a priority for future work is to extend this to the competing risk situation.

To conclude, through our simulation study and data application we have illustrated strengths and weaknesses of three distinct covariate adjustment approaches for treatment specific marginal cumulative incidence curves, emphasizing their behavior under model misspecification. With our work, we have bridged the gap between doubly robust covariate adjustment methods for standard survival curves and adjustment methods for cumulative incidence curves in competing risks settings. Future studies could further extend these methods to more complex multi-state settings.

## Acknowledgments

The authors thank the registration team of the Netherlands Comprehensive Cancer Organisation (IKNL) for the collection of data for the Netherlands Cancer Registry as well as IKNL staff for scientific advice.

## References

1. Saad ED. Endpoints in advanced breast cancer: Methodological aspects & clinical implications. *Indian J Med Res*. 2011.
2. Putter H, Fiocco M, Gekus RB. Tutorial in biostatistics: Competing risk and multi-state models. *Stat Med*. 2007. doi:10.1002/sim.2712
3. Nørgaard M, Ehrenstein V, Vandenbroucke JP. Confounding in observational studies based on large health care databases: Problems and potential solutions – a primer for the clinician. *Clin Epidemiol*. 2017. doi:10.2147/CLEP.S129879
4. Hernan MA, Robins JM. *Causal Inference: What If*. CRC Press; 2023. https://doi.org/10.1201/9781315374932.
5. Goetghebeur E, le Cessie S, De Stavola B, Moodie EEM, Waernbaum I. Formulating causal questions and principled statistical answers. *Stat Med*. 2020;39(30):4922-4948. doi:10.1002/sim.8741
6. Fine JP, Gray RJ. A Proportional Hazards Model for the Subdistribution of a Competing Risk. *J Am Stat Assoc*. 1999;94(446):496-509. doi:10.1080/01621459.1999.10474144
7. Cox DR, Oakes D. *Analysis of Survival Data*.; 1984. doi:10.1201/9781315137438
8. Austin PC, Lee DS, Fine JP. Introduction to the Analysis of Survival Data in the Presence of Competing Risks. *Circulation*. 2016;133(6):601-609. doi:10.1161/CIRCULATIONAHA.115.017719
9. Andersen PK, Syriopoulou E, Parner ET. Causal inference in survival analysis using pseudo-observations. *Stat Med*. 2017;36(17):2669-2681.
10. Young JG, Stensrud MJ, Tchetgen Tchetgen EJ, Hernán MA. A causal framework for classical statistical estimands in failure-time settings with competing events. *Stat Med*. 2020;39(8):1199-1236.
11. Tanaka S, Brookhart MA, Fine JP. G-estimation of structural nested mean models for competing risks data using pseudo-observations. *Biostatistics*. 2020;21(4):860-875.
12. Austin PC, Stuart EA. Moving towards best practice when using inverse probability of treatment weighting (IPTW) using the propensity score to estimate causal treatment effects in observational studies. *Stat Med*. 2015;34(28):3661-3679. doi:10.1002/sim.6607
13. Neumann A, Billionnet C. Covariate adjustment of cumulative incidence functions for competing risks data using inverse probability of treatment weighting. *Comput Methods Programs Biomed*. 2016;129:63-70. doi:10.1016/j.cmpb.2016.03.008
14. Choi S, Kim C, Zhong H, Ryu ES, Han SW. Adjusted-crude-incidence analysis of multiple treatments and unbalanced samples on competing risks. *Stat Interface*. 2019;12(3). doi:10.4310/19-SII560
15. Therneau TM, Crowson CS, Atkinson EJ. Adjusted survival curves. *R J*. 2015. https://cran.r-project.org/web/packages/survival/vignettes/adjcurve.pdf.
16. Kipourou DK, Charvat H, Rachet B, Belot A. Estimation of the adjusted cause-specific cumulative probability using flexible regression models for the cause-specific hazards. *Stat Med*. 2019;38(20):3896-3910. doi:10.1002/sim.8209
17. Hu ZH, Peter Gale R, Zhang MJ. Direct adjusted survival and cumulative incidence curves for observational studies. *Bone Marrow Transplant*. 2020;55(3):538-543. doi:10.1038/s41409-019-0552-y
18. Chatton A, Le Borgne F, Leyrat C, et al. G-computation, propensity score-based methods, and targeted maximum likelihood estimator for causal inference with different covariates sets: a comparative simulation study. *Sci Rep*. 2020;10(1):1-13. doi:10.1038/s41598-020-65917-x

19. Funk MJ, Westreich D, Wiesen C, Stürmer T, Brookhart MA, Davidian M. Doubly robust estimation of causal effects. *Am J Epidemiol*. 2011;173(7):761-767. doi:10.1093/aje/kwq439
20. Wang J. A simple, doubly robust, efficient estimator for survival functions using pseudo observations. *Pharm Stat*. 2018;17(1):38-48. doi:10.1002/pst.1834
21. Cole SR, Hernán MA. Adjusted survival curves with inverse probability weights. *Comput Methods Programs Biomed*. 2004. doi:10.1016/j.cmpb.2003.10.004
22. Austin PC. Variance estimation when using inverse probability of treatment weighting (IPTW) with survival analysis. *Stat Med*. 2016;35(30):5642-5655. doi:10.1002/sim.7084
23. Sjölander A. Regression standardization with the R package stdReg. *Eur J Epidemiol*. 2016;31(6):563–574. doi:10.1007/s10654-016-0157-3
24. Lunceford JK, Davidian M. Stratification and weighting via the propensity score in estimation of causal treatment effects: A comparative study. *Stat Med*. 2004;23(19):2937-2960. doi:10.1002/sim.1903
25. Robins JM, Rotnitzky A, Zhao LP. Estimation of Regression Coefficients When Some Regressors are not Always Observed. *J Am Stat Assoc*. 1994;89(427):846-866. doi:10.1080/01621459.1994.10476818
26. Denz R, Klaaßen-Mielke R, Timmesfeld N. A comparison of different methods to adjust survival curves for confounders. *Stat Med*. 2023;(January):1-19. doi:10.1002/sim.9681
27. Gerds TA. prodlim: Product-Limit Estimation for Censored Event History Analysis. R package version 2019.11.13. https://CRAN.R-project.org/package=prodlim. 2019.
28. Andersen PK, Pohar Perme M. Pseudo-observations in survival analysis. *Stat Methods Med Res*. 2010;19(1):71-99. doi:10.1177/0962280209105020
29. Bender R, Augustin T, Blettner M. Generating survival times to simulate Cox proportional hazards models. *Stat Med*. 2005;24(11):1713-1723. doi:10.1002/sim.2059
30. Piessens R, de Doncker-Kapenga E, Überhuber CW, Kahaner DK. *Quadpack: A Subroutine Package for Automatic Integration*. Vol 1. Berlin, Heidelberg: Springer Berlin / Heidelberg; 1983.
31. Therneau TM. A Package for Survival Analysis in R. *R Packag version 32-11*. 2021. https://cran.r-project.org/package=survival.
32. Ozenne B, Sørensen AL, Scheike T, Torp-Pedersen C, Gerds TA. riskRegression: Predicting the risk of an event using cox regression models. *R J*. 2020. doi:10.32614/rj-2017-062
33. Wickham H. *Ggplot2: Elegant Graphics for Data Analysis*. Springer-Verlag New York; 2016. https://ggplot2.tidyverse.org.
34. van Kesteren E-J. vankesteren/firatheme: firatheme version 0.2.4. April 2021. doi:10.5281/zenodo.4679413
35. Mersmann O. microbenchmark: Accurate Timing Functions. 2019. https://cran.r-project.org/package=microbenchmark.
36. Lumley T. *Complex Surveys: A Guide to Analysis Using R: A Guide to Analysis Using R*. John Wiley and Sons; 2010.
37. Yoshida K, Bartel A. tableone: Create “Table 1” to Describe Baseline Characteristics with or without Propensity Score Weights. 2022. https://cran.r-

project.org/package=tableone.
38. van Maaren MC, le Cessie S, Strobbe LJA, Groothuis-Oudshoorn CGM, Poortmans PMP, Siesling S. Different statistical techniques dealing with confounding in observational research: measuring the effect of breast-conserving therapy and mastectomy on survival. *J Cancer Res Clin Oncol*. 2019;145(6):1485-1493. doi:10.1007/s00432-019-02919-x
39. van Maaren MC, de Munck L, de Bock GH, et al. 10 year survival after breast-conserving surgery plus radiotherapy compared with mastectomy in early breast cancer in the Netherlands: a population-based study. *Lancet Oncol*. 2016. doi:10.1016/S1470-2045(16)30067-5
40. Rahman GA. Editorial: Breast conserving therapy: A surgical technique where little can mean more. *J Surg Tech Case Rep*. 2011. doi:10.4103/2006-8808.78459
41. Lee WQ, Tan VKM, Choo HMC, et al. Factors influencing patient decision-making between simple mastectomy and surgical alternatives. *BJS open*. 2019;3(1). doi:10.1002/bjs5.50105
42. Robins J, Sued M, Lei-Gomez Q, Rotnitzky A. Comment: Performance of Double-Robust Estimators When ``Inverse Probability'' Weights Are Highly Variable. *Stat Sci*. 2008;22. doi:10.1214/07-STS227D
43. Golinelli D, Ridgeway G, Rhoades H, Tucker J, Wenzel S. Bias and variance trade-offs when combining propensity score weighting and regression: With an application to HIV status and homeless men. *Heal Serv Outcomes Res Methodol*. 2012;12(2-3):104-118. doi:10.1007/s10742-012-0090-1
44. Binder N, Gerds TA, Andersen PK. Pseudo-observations for competing risks with covariate dependent censoring. *Lifetime Data Anal*. 2014;20(2):303-315.
45. Overgaard M, Parner ET, Pedersen J. Pseudo-observations under covariate-dependent censoring. *J Stat Plan Inference*. 2019;202:112-122.
46. Bai X, Tsiatis AA, O'Brien SM. Doubly-robust estimators of treatment-specific survival distributions in observational studies with stratified sampling. *Biometrics*. 2013;69(4):830-839. doi:10.1111/biom.12076
47. Wang X, Beste LA, Maier MM, Zhou X-H. Double robust estimator of average causal treatment effect for censored medical cost data. *Stat Med*. 2016;35(18):3101-3116. doi:10.1002/sim.6876
48. Hubbard AE, van der Laan MJ, Robins JM. Nonparametric locally efficient estimation of the Treatment Specific Survival distribution with right Censored Data and Covariates in Observational Studies. In: Halloran ME, Berry D, eds. *Statistical Models in Epidemiology, the Environment, and Clinical Trials*. New York, NY: Springer New York; 2000:135-177.
49. Ozenne BMH, Scheike TH, Stærk L, Gerds TA. On the estimation of average treatment effects with right-censored time to event outcome and competing risks. *Biometrical J*. 2020;62(3):751-763. doi:10.1002/bimj.201800298
50. Blanche PF, Holt A, Scheike T. On logistic regression with right censored data, with or without competing risks, and its use for estimating treatment effects. *Lifetime Data Anal*. 2023;29(2):441-482. doi:10.1007/s10985-022-09564-6
51. Gabriel EE, Sachs MC, Waernbaum I, et al. Propensity weighting plus adjustment in proportional hazards model is not doubly robust. 2024. https://arxiv.org/abs/2310.16207.